\documentclass[reprint,superscriptaddress,amsmath,amssymb,aps,showpacs,pra,]{revtex4-1}
\usepackage{appendix}
\usepackage{verbatim}
\usepackage{float}
\usepackage{color}
\usepackage{textcomp}
\usepackage{gensymb}
\usepackage{hyperref}
\hypersetup{
     colorlinks   = true,
     citecolor    = blue
}
\usepackage{siunitx}
\usepackage{graphicx}
\usepackage{dcolumn}
\usepackage{bm}
\usepackage{braket}
\usepackage{empheq}
\usepackage{nicefrac} 

\begin{document}

\title{Unconventional Thermal Magnon Hall Effect in a Ferromagnetic Topological Insulator}

\author{Christian Moulsdale}
\email{christian.moulsdale@postgrad.manchester.ac.uk}
\affiliation{\noindent Theoretical Physics Division, School of Physics and Astronomy, University of Manchester, Manchester M13 9PL, United Kingdom}%

\author{\noindent Pierre A. Pantale{\'o}n}
\affiliation{\noindent Theoretical Physics Division, School of Physics and Astronomy, University of Manchester, Manchester M13 9PL, United Kingdom}

\author{\noindent Ramon Carrillo\textendash Bastos}
\affiliation{Facultad de Ciencias, Universidad Aut{\'o}noma de Baja California,
Apartado Postal 1880, 22800, Ensenada, Baja California, M{\'e}xico}

\author{Yang Xian}
\affiliation{\noindent Theoretical Physics Division, School of Physics and Astronomy, University of Manchester, Manchester M13 9PL, United Kingdom}%

\date{\today}

\begin{abstract}
We present theoretically the thermal Hall effect of magnons in a ferromagnetic lattice with a Kekul\'e\textendash O coupling (KOC) modulation and a Dzyaloshinskii\textendash Moriya interaction (DMI). Through a strain-based mechanism for inducing the KOC modulation, we identify four topological phases in terms of the KOC parameter and DMI strength. We calculate the thermal magnon Hall conductivity ${\kappa^{xy}}$ at low temperature in each of these phases. We predict an unconventional conductivity due to a non-zero Berry curvature emerging from band proximity effects in the topologically trivial phase. We find sign changes of ${\kappa^{xy}}$ as a function of the model parameters, associated with the local Berry curvature and occupation probability of the bulk bands. Throughout, ${\kappa^{xy}}$ can be easily tuned with external parameters such as the magnetic field and temperature. 
\end{abstract}

\maketitle

\section{\label{sec:intro}Introduction}

\begin{figure*}[htp]
    \centering
    \includegraphics[width=\linewidth]{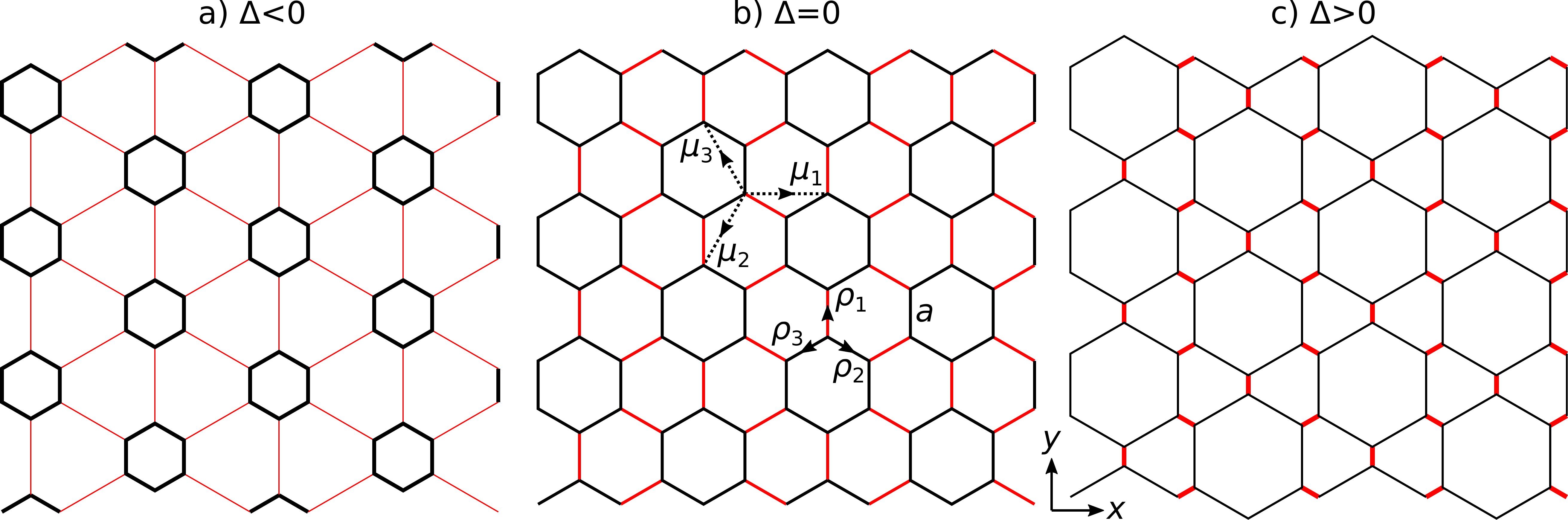}
    \caption{\small{} (Colour online) KOC modulation of the honeycomb lattice for different values of the KOC parameter $\Delta$. The intracell (intercell) couplings are shown as black (red) lines of widths proportional to their strength. The isotropic, unstrained case with ${\Delta=0}$ is shown in b), with the unstrained NN vectors ${\bm{\rho}}$, NNN vectors ${\bm{\mu}}$ and lattice parameter $a$ indicated.}
    \label{fig:kek}
\end{figure*}

The thermal Hall effect is an unconventional Hall effect where a heat current is found perpendicular to a temperature gradient in a material~\cite{Righi,Leduc}. In a ferromagnetic system, these heat currents are carried by magnons~\cite{Matsumoto2011,Matsumoto2011b,Matsumoto2011c,Murakami2017a}, which are weakly interacting quasiparticles obeying Bose\textendash Einstein statistics~\cite{Holstein1940}. Unlike the Hall effect of electrons~\cite{10.2307/2369245}, this is not a result of the Lorentz force as magnons carry no electronic charge. Then, as neutral quasiparticles, they can propagate over large distances without dissipation by Joule heating~\cite{Kajiwara2010}. The thermal Hall effect of magnons has been observed in pyrochlore ferromagnetic insulators~\cite{Onose2010,Ideue2012}, yttrium iron garnets~\cite{Madon2014,Tanabe2016}, a kagom{\'e} magnet~\cite{Hirschberger2015} and a frustrated pryocholore quantum magnet~\cite{Hirschberger2015a}.

The emergence of any Hall effect is typically a consequence of time reversal symmetry breaking which induces topologically non-trivial band gaps~\cite{Murakami2003,Sinova2004,Day2008}. This can be achieved in an electronic system in many ways including with an external magnetic field~\cite{Thouless1982} or by introducing spin-orbit coupling (SOC)~\cite{Sinova2004,Kane2005}. Likewise, the Dzyaloshinskii\textendash Moriya interaction (DMI) representing SOC in magnetic systems~\cite{Dzyaloshinsky1958,Moriya1960} induces a non-trivial band topology~\cite{Katsura2010}. On the other hand, a non-trivial band topology is found in the absence of DMI in frustrated antiferromagnets~\cite{Owerre2017} or by introducing the magnetic dipolar interaction~\cite{Matsumoto2011,Matsumoto2011b}. From the bulk Hamiltonian and when a nontrivial gap is present, we can use the bulk-boundary correspondence~\cite{PhysRevB.84.125132} to predict the existence of robust magnon edge currents~\cite{PhysRevB.84.195452,PhysRevB.95.235118} which are independent of the geometry of the system~\cite{Matsumoto2011}. The edge states of an isotropic 2D honeycomb ferromagnet with DMI have been studied theoretically for a system with zigzag~\cite{Pantaleon2017, Pantaleon2018a}, armchair~\cite{Pantaleon2017a} and bearded boundaries~\cite{Pantaleon2018a}. These may be observed experimentally in monolayers of the ferromagnetic material chromium triiodide ${(\mathrm{CrI}_3)}$, which has been shown to possess large intrinsic DMI and consequently a non-trivial band topology~\cite{Chen2018}.

In graphene, bond modulations can be induced by local changes in the position of the carbon atoms due to the absorption of adatoms on its surface~\cite{Cheianov2009} or by a proximity effect~\cite{Lin2017,Gutierrez2016}. One such example is the KOC modulation, depicted in Fig.~\ref{fig:kek}, which is an intrinsic instability of carbon nanotubes~\cite{Chamon2000} and graphene~\cite{Ryu2009a,Weeks2010,Garcia-Martinez2013}. In this case, a non-uniform strain field results in bonds of different strength within and between hexagonal unit cells similar to benzene molecules~\cite{Gamayun2018}. Moreover, the adsorption of lithium adatoms on the surface of ${\mathrm{CrI}_3}$ has been predicted to result in an enhanced ferromagnetism and an increased Curie temperature~\cite{Guo2018}. Therefore, it is interesting to ask if the KOC modulation can be achieved in its ferromagnetic couplings through the selective adsorption of such adatoms.

In this paper, we report by a theoretical investigation that the thermal Hall effect of magnons is found in a KOC modulated honeycomb ferromagnet with DMI. We extend our previous model~\cite{Pantaleon2018b} to include a strain-based mechanism and NNN ferromagnetic exchange couplings, resulting in four topological phases. We find a small but non-vanishing thermal Hall conductivity in the trivial phase despite the lack of topologically protected edge states. Furthermore, unlike the isotropic model without the KOC modulation, we find sign changes of the thermal Hall conductivity with respect to the various model parameters. These are explained in terms of the competing contributions of the bulk bands. The model studied here may be useful for the future design of two\textendash dimensional thermal components since the thermal Hall conductivity can be easily tuned with external parameters.

This paper is structured as follows. In Sec.~\ref{sec:model}, we introduce the Heisenberg model with KOC modulation and DMI. In Sec.~\ref{sec:thermal}, we give expressions for the Chern numbers and thermal Hall conductivity. In Sec.~\ref{sec:results}, we discuss the properties of the four topological phases found before considering the effects of the magnetic field. Finally, Sec.~\ref{sec:conc} is devoted to conclusions.

\section{\label{sec:model}Bond Modulated Heisenberg Model}

Recently, it has been shown that different topological phases can be induced in a ferromagnetic honeycomb lattice by introducing the KOC modulation and DMI~\cite{Pantaleon2018b}. In this paper, we provide evidence of the thermal Hall effect of magnetic spin excitations in a similar model, whose Hamiltonian is given by
\begin{equation}
\begin{split}
H = &-\sum_{\left\langle ij \right\rangle} J_{ij} \bm{S}_i \cdot \bm{S}_j -\sum_{\left\langle\left\langle ij \right\rangle\right\rangle} J'_{ij} \bm{S}_i \cdot \bm{S}_j -A \sum_i (S_i^z)^2 \\
&+ \sum_{\left\langle\left\langle ij \right\rangle\right\rangle} \bm{D}_{ij} \cdot (\bm{S}_i \times \bm{S}_j)-g\mu_BB\sum_i S_i^z,
\label{eq:H}
\end{split}
\end{equation}
in terms of the spin operators ${\bm{S}_i}$. The first two terms in the above equation represent nearest-neighbour (NN) and next-nearest-neighbour (NNN) ferromagnetic couplings respectively. The third term represents an easy-axis anisotropy in the ferromagnetic coupling with the $z$-axis identified as the easy-axis. The fourth term represents the NNN antisymmetric exchange (DMI), where $\bm{D}_{ij}$ is the DMI vector whose orientation depends on the lattice geometry and follows the rules set out by Moriya~\cite{Moriya1960}. The last term represents a Zeeman coupling with an external magnetic field ${\bm{B}=B\bm{e}_z}$, where $g$ is the gyromagnetic ratio and ${\mu_B=e\hbar/2m_e}$ is the Bohr magneton.

\begin{figure}[t]
    \begin{centering}
        \includegraphics[width=\linewidth]{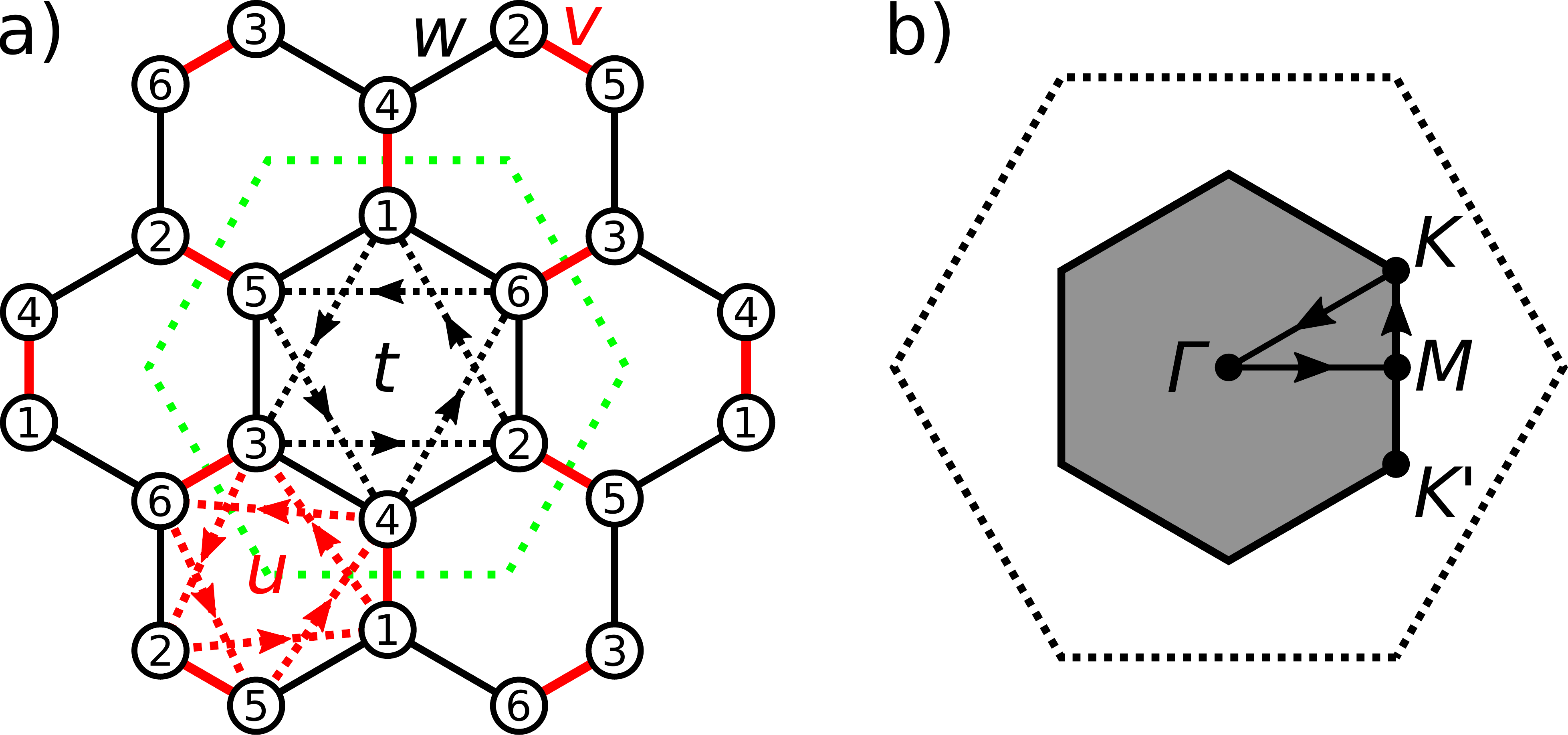}
    \par\end{centering}
    \caption{{\small{} (Colour online) a) The KOC modulated honeycomb lattice ${(\Delta>0)}$ with DMI complete with the indexed site basis in the unit cell shown as a dotted green hexagon. The strengths of the various couplings are indicated and described in the main text. The arrows give the directions in which ${\nu_{ij} = 1}$. b) The vertices of the unstrained Brillouin zone (dashed hexagon) are folded into the center of the strained Brillouin zone (shaded hexagon) due to the KOC modulation. The path $\Gamma MK \Gamma$ in $\bm{k}$-space between the high symmetry points shown as black circles used in later figures is shown in b). \label{fig:lat}}}
\end{figure}

To generate the KOC modulation shown in Fig.~\ref{fig:kek}, we introduce a non-uniform strain field~\cite{Ferreiros2018} controlled by the KOC parameter $\Delta$. Following the convention of Gamayun et al.~\cite{Gamayun2018}, the NN coupling amplitudes within a unit cell (intracell) and between unit cells (intercell) shown in Fig.~\ref{fig:lat}a) are given by
\begin{equation}
\begin{split}
    w &= (1-\Delta)J, \\
    v &= (1+2\Delta)J,
\label{eq:wv}
\end{split}
\end{equation}
respectively, where the unstrained, isotropic NN coupling amplitude is given by ${J_{ij}=J}$.  The intracell bonds are strained by $\delta$, so that the the spin-spin distance is ${(1+\delta)a}$ and the exchange coupling becomes ${w = J e^{-\beta \delta}}$, like in graphene~\cite{Vozmediano2010,Guinea2012,Andrade2019}. The Gruneissen\textendash like parameter $\beta$ describes the response of the couplings to the local strain and, similarly to graphene \cite{Mohiuddin2009,Cheng2011}, depends on the microscopic properties of the magnetic system. To linear order in the strain, we write
\begin{equation}
    \label{eq:strain}
    w \simeq (1-\beta \delta)J,
\end{equation}
with similar expressions found for the other couplings. Comparing this to Eq.~(\ref{eq:wv}), we find that the intracell strain is ${\delta=\nicefrac{\Delta}{\beta}}$ while the intercell strain is ${-2\delta}$. This convention preserves the lattice vectors and hence the size of the unit cell as shown in Fig.~\ref{fig:lat}a).

We now consider the modulation of the NNN couplings. Since the lattice sites lie in the $xy$-plane, the DMI vector takes the form ${\bm{D}_{ij} = D_{ij} \nu_{ij} \bm{e}_z}$, where $D_{ij}$ is the DMI strength and ${\nu_{ij}= 1(-1)}$ for counterclockwise (clockwise) couplings~\cite{Kane2005}. The unstrained values of $J'_{ij}$ and $D_{ij}$ are $J'$ and $D$ respectively. As shown in Fig.~\ref{fig:lat}a), there are two distinct kinds of NNN couplings: intracell ${(t)}$ with a strain of $\delta$ and intercell ${(u)}$ with a strain of ${-\nicefrac{\delta}{2}}$. We assume that the NNN couplings have the same Gruneissen\textendash like parameter, as in graphene~\cite{Botello-Méndez2018}. Thus, Eq.~(\ref{eq:strain}) gives
\begin{equation}
\begin{split}
t &= (1-\Delta) \sqrt{J'^2+D^2}, \\
u &= (1+\nicefrac{\Delta}{2}) \sqrt{J'^2+D^2},
\label{eq:tur}
\end{split}
\end{equation}
as the magnitudes of their corresponding NNN coupling amplitudes to linear order in $\Delta$. The ansatz in our previous work~\cite{Pantaleon2018b} corresponds to the NNN Gruneissen\textendash like parameter being ${2\beta}$.

The KOC modulated Heisenberg model consisting of the first two terms in Eq.~(\ref{eq:H}) features an SO(3) rotational symmetry of the magnetization vector. The Zeeman term totally breaks this symmetry, favouring the alignment of the magnetization with the external magnetic field $\bm{B}$. Thus, the mean-field ground state energy of our model with ${\bm{S}_i = \xi S \bm{e}_z}$ is ${E_0 = -\frac{1}{2}N(3J+6J'+2A)S^2-Ng \mu_B |B|}$, where ${\xi = \mathrm{sgn}(B)}$ and $N$ is the total number of lattice points. This is independent of both the KOC modulation and the DMI.

We perform Holstein\textendash Primakoff (HP) transformations on the spin operators~\cite{Holstein1940}, expressing them in terms of boson operators ${a_i(a_i^\dag)}$ which destroy (create) a magnon at sublattice point $i$. Each magnon carries a quantum $\hbar$ of angular momentum. We make the linear spin-wave approximation and thus neglect the higher-order terms in the boson operators which result in magnon-magnon interactions. The linear HP transformations are given by
\begin{equation}
    \label{eq:HP}
    S_i^+ = \sqrt{2S} a_i, \quad S_i^- = \sqrt{2S} a_i^\dag, \quad S_i^z = S-a_i^\dag a_i,
\end{equation}
where the spin ladder operators are ${S_i^\pm = S_i^x \pm i S_i^y}$. This corresponds to a ferromagnetic ground state aligned in the positive $z$-direction, appropriate for ${B>0~(\xi=1)}$, with the magnetization ${\left\langle S^z \right\rangle=S-\left\langle a^\dag a \right\rangle}$ identified as the order parameter. The spin-flipped HP transformations appropriate for ${B<0~(\xi=-1)}$ are given by Eq.~(\ref{eq:HP}) with ${S_i^+ \leftrightarrow S_i^-}$ and ${S_i^z \rightarrow -S_i^z}$. This flips the sign of the DMI term in Eq.~(\ref{eq:H}), so we multiply the DMI strengths by $\xi$ to make this term invariant. Hence, magnons propagating along NNN couplings gain a phase ${\nu_{ij}\phi}$, where ${\phi=\xi \arctan(\nicefrac{D}{J'})}$.

Finally, we perform a Fourier transformation on the boson operators to move into ${\bm{k}}$-space. Then, the Hamiltonian in Eq.~(\ref{eq:H}) is written as
\begin{equation}
\label{eq:HM}
H = E_0 + \sum_{\bm{k}} \Psi_{\bm{k}}^\dag M_{\bm{k}} \Psi_{\bm{k}},
\end{equation}
where the summation is across all states of wavevector $\bm{k}$ in the Brillouin zone (BZ), shown as a filled hexagon in Fig. \ref{fig:lat}b). Compared to the BZ of the isotropic 2-band model, shown as a dashed hexagon in Fig. \ref{fig:lat}b), this is smaller in area by a factor of 3 and rotated by $30\degree$ around the zone center $\Gamma$. In Eq.~(\ref{eq:HM}), ${M_{\bm{k}}}$ is a 6x6 matrix in the basis ${\Psi_{\bm{k}}^\dag = (a_{1\bm{k}}^\dag, a_{2\bm{k}}^\dag, \cdots, a_{6\bm{k}}^\dag)}$ shown in Fig.~\ref{fig:lat}a) given by 
\begin{equation}
    \label{eq:M}
    M_{\bm{k}} = \epsilon_0 I_6 + \begin{pmatrix}
        M_{AA} & M_{AB} \\
        M_{BA} & M_{BB}
    \end{pmatrix}.
\end{equation}
The first term represents the magnon on-site potential ${\epsilon_0=3(J+2J')S+h}$, where the effective magnetic field strength is ${h=2A S+g\mu_B|B|}$ and $I_6$ is the 6x6 identity matrix. The off-diagonal matrices in the second term represent the NN coupling and are given by
\begin{equation}
    \label{eq:MAB}
    M_{AB}=M_{BA}^* = -S
    \begin{pmatrix}
        \gamma_{1v} & \gamma_{3w} & \gamma_{2w} \\
        \gamma_{3w} & \gamma_{2v} & \gamma_{1w} \\
        \gamma_{2w} & \gamma_{1w} & \gamma_{3v}
    \end{pmatrix}.
\end{equation}
The intracell and intercell gamma factors are given respectively by
\begin{equation}
    \gamma_{iw}=we^{i\bm{k}\cdot(1+\delta)\bm{\rho}_i}, \quad \gamma_{iv}=ve^{i\bm{k}\cdot(1-2\delta)\bm{\rho}_i},
\end{equation}
in terms of the unstrained NN vectors ${\bm{\rho} = \{}{(0, 1),~(\nicefrac{\sqrt{3}}{2}, -\nicefrac{1}{2}),~(-\nicefrac{\sqrt{3}}{2}, -\nicefrac{1}{2})\}a}$ shown in Fig.~\ref{fig:kek}b), where $a$ is the unstrained lattice constant. Similarly, the diagonal components representing the NNN coupling are given by
\begin{equation}
    \label{eq:MAA}
    M_{AA} = M_{BB}|_{d_i \leftrightarrow d_i^*} = -S
    \begin{pmatrix}
        0 & z^*d_3^* & zd_2 \\
        zd_3 & 0 & z^*d_1^* \\
        z^*d_2^* & zd_1 & 0
    \end{pmatrix}.
\end{equation}
The complex number is ${z=e^{-i\phi}}$ and
\begin{equation}
    \label{eq:d}
    d_i = \eta_i^{\delta} (t \eta_i + u \eta_{i+1} + u \eta_{i-1}),
\end{equation}
with the indices defined modulus 3. In Eq.~(\ref{eq:d}), ${\eta_i = e^{i\bm{k}\cdot\bm{\mu}_i}}$ in terms of the unstrained NNN vectors ${\bm{\mu}_i = \bm{\rho}_{i+1}-\bm{\rho}_{i-1}}$ which are also shown in Fig.~\ref{fig:kek}b).

We solve the time-independent Schr\"odinger equation (TISE) ${M_{\bm{k}} \ket{\psi_{\lambda\bm{k}}} = \epsilon_{\lambda\bm{k}} \ket{\psi_{\lambda\bm{k}}}}$ to find the corresponding energy eigenvalue ${\epsilon_{\lambda\bm{k}}}$ and eigenvector ${\ket{\psi_{\lambda\bm{k}}}}$ of a band $\lambda$. These bands ${\lambda = 1, 2, \cdots, 6}$ are indexed by increasing energy. We proceed numerically for general values of $\bm{k}$, although the TISE is exactly solvable at the high symmetry points $\Gamma$, $K$ and $M$ of the BZ shown in Fig.~\ref{fig:lat}b).

\section{\label{sec:thermal}Thermal Magnon Hall Effect}

\subsection{Berry Phase and Chern numbers}

\begin{figure}[t]
    \includegraphics[width=\linewidth]{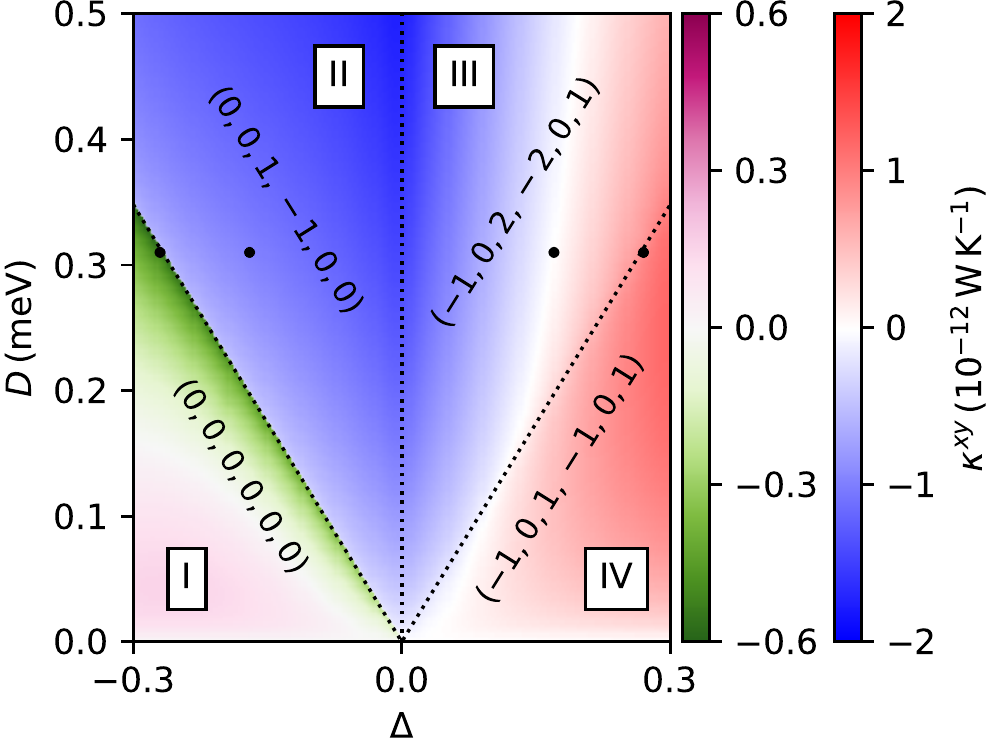}
    \caption{\small{}(Colour online) The thermal Hall conductivity $\kappa^{xy}$ of a KOC modulated ${\mathrm{CrI}_3}$ monolayer against the KOC parameter $\Delta$ and DMI strength $D$ at temperature ${T=\SI{40}{K}}$ with magnetic field ${B=0^+ \, \mathrm{T} \, (\xi=1)}$. Each topological phase is distinguished by its set of Chern numbers ${(C_1, C_2, \cdots, C_6)}$. Gap-closing transitions separating phases are depicted as dotted black lines. The four points in each phase we use in later figures lying on the line ${D=\SI{0.31}{meV}}$ with ${\Delta = -0.27}$ (I), ${-0.17}$ (II), ${0.17}$ (III) and ${0.27}$ (IV) are shown as black dots.
    \label{fig:kTphase}}
\end{figure}

Nonzero DMI is a consequence of inversion symmetry breaking and breaks the time reversal symmetry of the component of the magnetization parallel to the DMI vector~\cite{Kim2016a, Owerre2016d}. Then, in a honeycomb ferromagnetic lattice, magnons accumulate an additional phase ${\nu_{ij}\phi}$ upon propagation between NNN sites and a nontrivial band topology arises characterised by a nonzero Berry curvature~\cite{berry84}. In 2D lattice systems, the Berry curvature of a band indexed by $\lambda$ is given by
\begin{equation}
\label{eq:Omega}
    \Omega_{\lambda \bm{k}} = -2 \sum_{\lambda' \neq \lambda} \mathrm{Im} \frac{\braket{\psi_{\lambda \bm{k}}|\partial_{k_x} M_{\bm{k}}|\psi_{\lambda' \bm{k}}} \braket{\psi_{\lambda' \bm{k}}|\partial_{k_y} M_{\bm{k}}|\psi_{\lambda \bm{k}}}}{(\epsilon_{\lambda \bm{k}} - \epsilon_{\lambda' \bm{k}})^2}.
\end{equation}
The Chern number ${C_\lambda}$ of the band $\lambda$ is given by the integral of its Berry curvature about the BZ:
\begin{equation}
\label{eq:C}
C_\lambda = \frac{1}{2 \pi} \int_\mathrm{BZ} d^2 k \Omega_{\lambda \bm{k}}.
\end{equation}
We use the algorithm of Fukui~et~al.~\cite{Fukui2005} to calculate the Chern numbers. In our six-band system, each topological phase is characterised by the set of Chern numbers of the bulk bands ${(C_1, C_2, \cdots, C_6)}$, as shown in Fig. \ref{fig:kTphase}. Another relevant quantity is the winding number $\nu_\zeta$ of the topologically protected edge states in band gap ${\zeta=1,\cdots,5}$ between bulk bands $\zeta$ and $\zeta+1$, given by \cite{Shindou2013,Mook2014}
\begin{equation}
\label{eq:wind}
\nu_{\zeta} = \sum_{\lambda \leq \zeta} C_{\lambda}.
\end{equation}
We find a number ${|\nu_\zeta|}$ of topological edge states traversing this band gap, propagating (counter-)clockwise around the system if ${\mathrm{sgn} (\nu_\zeta)=(-)1}$.

\subsection{Thermal Hall conductivity}

A direct consequence of a nonzero Berry curvature in a magnonic system is the presence of thermal currents perpendicular to an applied temperature gradient, described by the thermal Hall conductivity $\kappa^{xy}$. In a series of seminal works~\cite{Matsumoto2011,Matsumoto2011b,Matsumoto2011c}, Matsumoto and Murakami demonstrated that, at low temperature $T$, $\kappa^{xy}$ can be split into two contributions, ${\kappa^{xy}=\kappa_E^{xy}+\kappa_O^{xy}}$, where $\kappa_E^{xy}$ is the contribution from the magnons' current density, given by
\begin{equation}
\begin{split}
\kappa_E^{xy} = &-\frac{1}{2\hbar T} \sum_{\lambda=1}^6 \int_\mathrm{BZ} \frac{d^2k}{(2\pi)^2} n_{\lambda\bm{k}}(T) \\
&\times \mathrm{Im} \braket{\partial_{k_x} \psi_{\lambda\bm{k}}|(M_{\bm{k}}+\epsilon_{\lambda\bm{k}}I_6)^2|\partial_{k_y} \psi_{\lambda\bm{k}}},
\label{eq:kxyE}
\end{split}
\end{equation}
and ${\kappa_O^{xy}}$ is the contribution from their reduced orbital momentum. The full expression for ${\kappa^{xy}}$ is given by
\begin{equation}
\label{eq:kxy}
\kappa^{xy} = -\frac{k_B^2T}{\hbar} \sum_{\lambda = 1}^6 \int_\mathrm{BZ} \frac{d^2k}{(2\pi)^2} c_2 [n_{\lambda\bm{k}}(T)]  \Omega_{\lambda\bm{k}}.
\end{equation}
Magnons obey Bose\textendash Einstein statistics so that their occupation function is given by ${n_{\lambda\bm{k}}(T) = (e^{\epsilon_{\lambda\bm{k}} / k_B T} - 1)^{-1}}$. The $c_2$ function is given in terms of this by ${c_2 (n)~=~(1 + n) \big[ \ln \big( 1 + \nicefrac{1}{n} \big) \big]^2 ~-~(\ln n)^2 ~-~ 2 \mathrm{Li}_2(-n)}$, where ${\mathrm{Li}_s(z)=\sum_{r=1}^\infty z^r/r^s}$ is a polylogarithm.

We can identify the contributions to the thermal Hall conductivity ${\kappa^{xy}_\lambda}$ of each band $\lambda$ in Eq.~(\ref{eq:kxy}) with ${\kappa^{xy} = \sum_{\lambda = 1}^6 \kappa^{xy}_\lambda}$. Both the Chern number $C_\lambda$ in Eq.~(\ref{eq:C}) and ${\kappa^{xy}_\lambda}$ of a band are integrals weighted by its Berry curvature ${\Omega_{\lambda\bm{k}}}$ and are intrinsically related. As a result, a band with a nonzero Chern number $C_\lambda$ will generally have a contribution ${\kappa^{xy}_\lambda}$ of opposite sign. Equally, a band gap with a positive winding number indicates the presence of clockwise-propagating edge states which give a negative contribution to ${\kappa^{xy}}$ and vice versa. At low temperatures, the lowest energy states have a greater occupation and so the dominant contribution comes from the lowest bands due to the $c_2$ function. As the temperature is increased, bands of increasing energy become occupied and provide significant contributions to ${\kappa^{xy}}$.

\begin{figure*}[ht!]
    \centering
    \includegraphics[width=\linewidth]{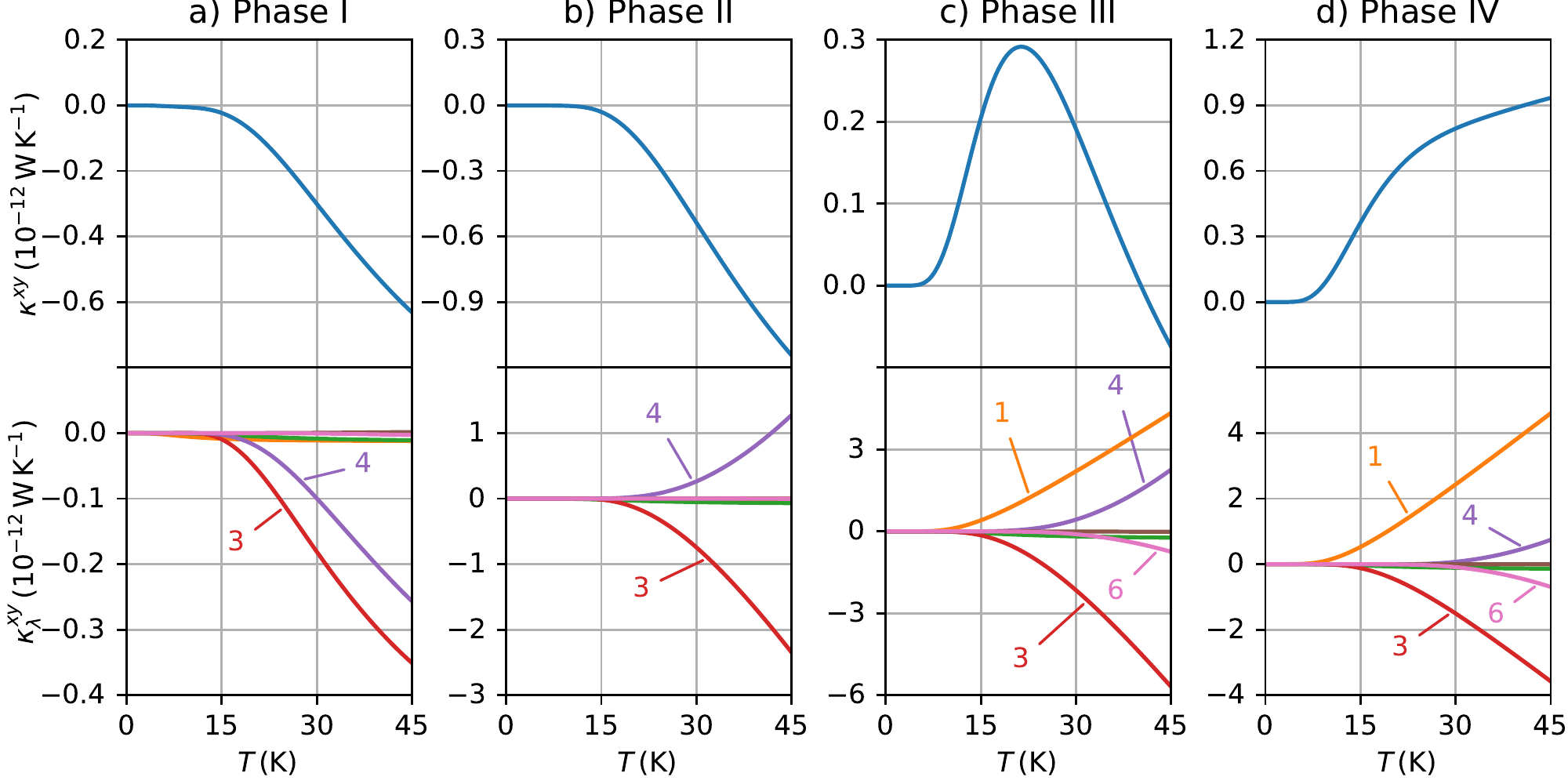}
    \caption{\small{} (Colour online) The thermal Hall conductivity (${\kappa^{xy}}$, top) and its band contributions (${\kappa^{xy}_\lambda}$, bottom) in a KOC modulated honeycomb lattice with DMI as a function of temperature $T$ for the points in each phase given in Fig.~\ref{fig:kTphase}.}
    \label{fig:kT}
\end{figure*}

\section{\label{sec:results}Results and discussion}

In the following calculations, we consider the parameters of a monolayer of the ferromagnetic material chromium triiodide ${(\mathrm{CrI}_3)}$. This is one the of chromium trihalides, which are a family of 2D ferromagnetic spin-$\nicefrac{3}{2}$ materials with gyromagnetic ratio ${g=3}$~\cite{Dillon1966,Wang2011,Zhang2015}. Using results from powder neutron diffraction~\cite{Chen2018}, the NN and NNN ferromagnetic coupling strengths are ${J=\SI{2.01}{meV}}$ and ${J'=\SI{0.16}{meV}}$ respectively. This model also considered next-next-nearest neighbour and interlayer couplings, but these have been neglected for being too weak and irrelevant to a monolayer, respectively. ${\mathrm{CrI}_3}$ has the highest Curie temperature ${T_c = \SI{45}{K}}$ of the chromium trihalides due to its large easy-axis anisotropy ${A=\SI{0.22}{meV}}$. It also has the largest intrinsic DMI strength ${D=\SI{0.31}{meV}}$ due to its heavy iodine anions, making it ideal for our theoretical analysis. Its Grunessein-like parameter has not yet been calculated, so we take ${\beta \simeq 2}$ as in graphene~\cite{Cheng2011} for demonstration purposes. This is reasonable since there is little variation of ${\kappa^{xy}}$ in the range ${1 \leq \beta \leq 4}$. For example, if the KOC parameter has a value of  ${\Delta=0.1}$, then the intracell bonds are stretched by 5\% and the intercell bonds compressed by 10\%.

The resulting phase diagram as a function of the KOC parameter $\Delta$ and the DMI strength $D$ is depicted in Fig.~\ref{fig:kTphase}, giving a total of four topological phases. The dotted black lines represent the critical regions where phase transitions occur. The v-shaped line gives the gap-closing condition at the BZ center $\Gamma$ between bulk bands 3 and 4 which changes the Chern numbers of these bands. The vertical line at ${\Delta=0}$ gives the gap-closing transition at the $K$ points which changes the Chern numbers of all bands except 2 and 5. This gap is trivial for ${\Delta<0}$ and non-trivial for ${\Delta>0}$. For small values of the NNN ferromagnetic coupling $J'$, the Chern numbers of each phase are the same as in our previous work~\cite{Pantaleon2018b}. Outside of the region considered in Fig.~\ref{fig:kTphase} where the couplings are linear in $\Delta$, we find the emergence of new topological phases.

The thermal magnon Hall conductivity ${\kappa^{xy}}$ at temperature ${T = \SI{40}{K}}$ is shown as a function of $\Delta$ and $D$ in Fig.~\ref{fig:kTphase}, revealing sign changes with respect to these parameters in phases I, III and IV. These sign changes are not observed in the absence of the KOC modulation~\cite{Owerre2016c}. Changing $T$ modifies the shape of the white region in Fig. \ref{fig:kTphase} where these sign changes are found. In a Kagome lattice, the sign of $\kappa^{xy}$ has been shown to depend on the winding number and the occupation probability of the edge magnons~\cite{Mook2014,Mook2014a}. Similar results are found in our system except for in phase~I, where all six Chern numbers are vanishing and so all five winding numbers from Eq.~(\ref{eq:wind}) are zero. Despite the resulting lack of topologically protected edge states, a nonzero ${\kappa^{xy}}$ is found in this phase.

In Fig.~\ref{fig:kT}, ${\kappa^{xy}}$ is depicted as a function of temperature $T$ for four different values of $\Delta$, each marked with a dot in Fig.~\ref{fig:kTphase}, which put the system into each of the topological phases. The total conductivity ${\kappa^{xy}}$ is shown in the top row and its corresponding band contributions ${\kappa^{xy}_\lambda}$ in the bottom row in each case. For phases II-IV, the only significant contributions come from the bands with non-vanishing Chern numbers. For the given parameters in phase~I, bands 3 and 4 are of particular significance.

In the following sections we discuss the different phases separately.

\begin{figure*}[ht]
\begin{centering}
\includegraphics[width=0.8\linewidth]{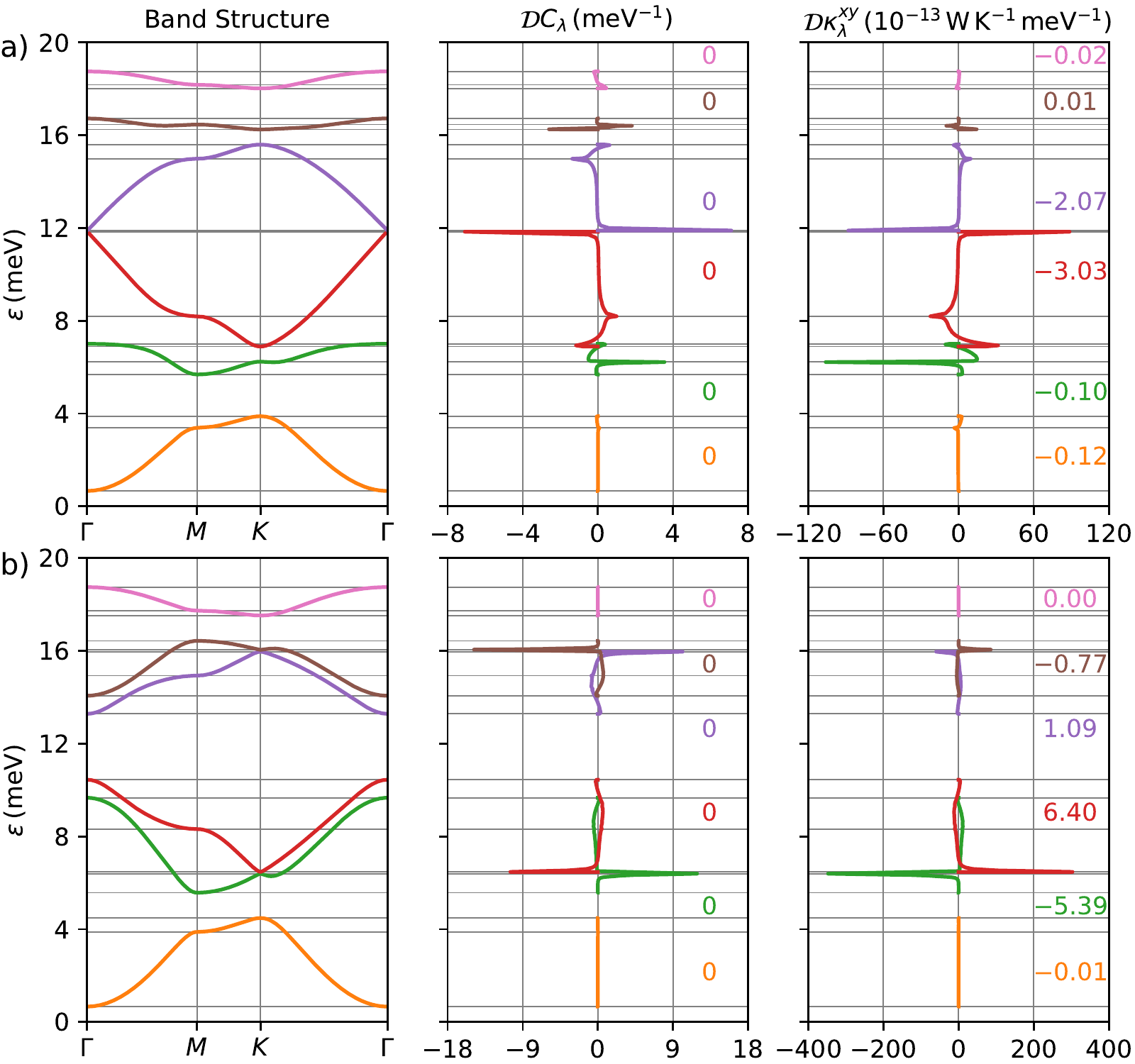}
\par\end{centering}
\caption{{\small{(Colour online) The band structure along the path $\Gamma MK \Gamma$ shown in Fig.~\ref{fig:lat} (left), the isoenergy surfaces of the Chern numbers ${\mathcal{D} C_\lambda}$ (middle) and the isoenergy surfaces of the contributions of each band to the thermal Hall conductivity ${\mathcal{D} \kappa^{xy}_\lambda}$} (right). The parameters are given in Fig.~\ref{fig:kTphase} with a) ${\Delta=-0.27}$ and ${D=\SI{0.31}{meV}}$ and b) ${\Delta=-0.2}$ and ${D=\SI{0.05}{meV}}$. The energies of the bands at the $\Gamma$, $M$ and $K$ points are highlighted in each plot by grey horizontal lines. The Chern numbers $C_\lambda$ and contribution to the conductivity ${\kappa^{xy}_\lambda}$ in units of ${\SI{e-13}{\watt\per\kelvin}}$ are given for each band as text of the corresponding colour in each subfigure. The colour scheme of the energy bands and its contributions are the same as in Fig.~\ref{fig:kT}.}\label{fig:iso}}
\end{figure*}

\subsection{\label{ssec:I}Phase I}

We first consider phase~I where, as shown in Fig.~\ref{fig:kTphase} and \ref{fig:kT}a), there is a nonzero thermal Hall conductivity ${\kappa^{xy}}$ despite this phase being topologically trivial. In addition, a change of sign of ${\kappa^{xy}}$ with respect to the KOC parameter $\Delta$ and the DMI strength $D$ can also be observed. To elucidate this, we write the contribution ${\mathcal{D}f_\lambda (\epsilon)}$ of the isonenergy surface of energy $\epsilon$ to an integral ${f_\lambda \equiv \int_\mathrm{BZ} d^2k F_{\lambda \bm{k}}}$ as~\cite{Mook2014a}
\begin{equation}
    \label{eq:iso}
    \mathcal{D}f_\lambda (\epsilon) = \int_\mathrm{BZ} d^2k \, \delta(\epsilon_{\lambda \bm{k}}-\epsilon) F_{\lambda \bm{k}},
\end{equation}
so that ${f_\lambda = \int_0^\infty d\epsilon \mathcal{D}f_\lambda(\epsilon)}$. The band structures, isoenergy surface contributions to the Chern numbers, ${\mathcal{D}C_\lambda}$, and corresponding contributions to the thermal conductivity, ${\mathcal{D} \kappa^{xy}_\lambda}$, are depicted in Fig.~\ref{fig:iso} for two points in phase~I of Fig. \ref{fig:kTphase} at ${T = \SI{40}{K}}$. The first point in Fig.~\ref{fig:iso}a) with ${\Delta=-0.27}$ and ${D=\SI{0.31}{meV}}$ is close to the gap-closing transition at $\Gamma$ with phase~II, so we find a pair of similarly-sized peaks in ${\mathcal{D}C_\lambda}$ of opposite sign at $\Gamma$ in bands 3 and 4 due to their proximity. In each of the bands, the contributions of the different isoenergy surfaces cancel, so that all the Chern numbers are vanishing. The integrand of our expression for ${\kappa_\lambda^{xy}}$ from Eq.~(\ref{eq:kxy}) contains the $c_2$ function, which suppresses the contributions ${\mathcal{D} \kappa^{xy}_\lambda}$ of higher isonenergy surfaces, allowing the ${\kappa_\lambda^{xy}}$ to take nonzero values. Bands 5 and 6 are so strongly suppressed as to have negligible contributions since the average thermal energy is ${k_BT = \SI{3.4}{meV}}$, lying within band 1. A large negative contribution is found around $\Gamma$ between bands 3 and 4 with a smaller negative contribution around the $K$ points between bands 2 and 3. Thus, we find a nonzero thermal conductivity ${\kappa^{xy}=\SI{-5.3e-13}{\watt\per\kelvin}}$ at this point, albeit significantly smaller than in neighbouring phases.

As we move away from phase~II in parameter space, the trivial gap between bands 3 and 4 at $\Gamma$ is opened up, while the gap between bands 2 and 3 (and 4 and 5) at the $K$ points is closed. This increases the contribution of the isoenergy surfaces at the $K$ points while suppressing those at $\Gamma$, resulting in a transition to positive ${\kappa^{xy}}$, as observed with ${\kappa^{xy}=\SI{1.3e-13}{\watt\per\kelvin}}$ at ${\Delta=-0.2}$, ${D=\SI{0.05}{meV}}$ in Fig.~\ref{fig:iso}b). Thus, the competition between the contributions of the $\Gamma$ and $K$ points in the parameter space results in the observed sign change of ${\kappa^{xy}}$ for a fixed temperature in Fig.~\ref{fig:kTphase}.

\begin{figure}[t]
    \centering
    \includegraphics[width=\linewidth]{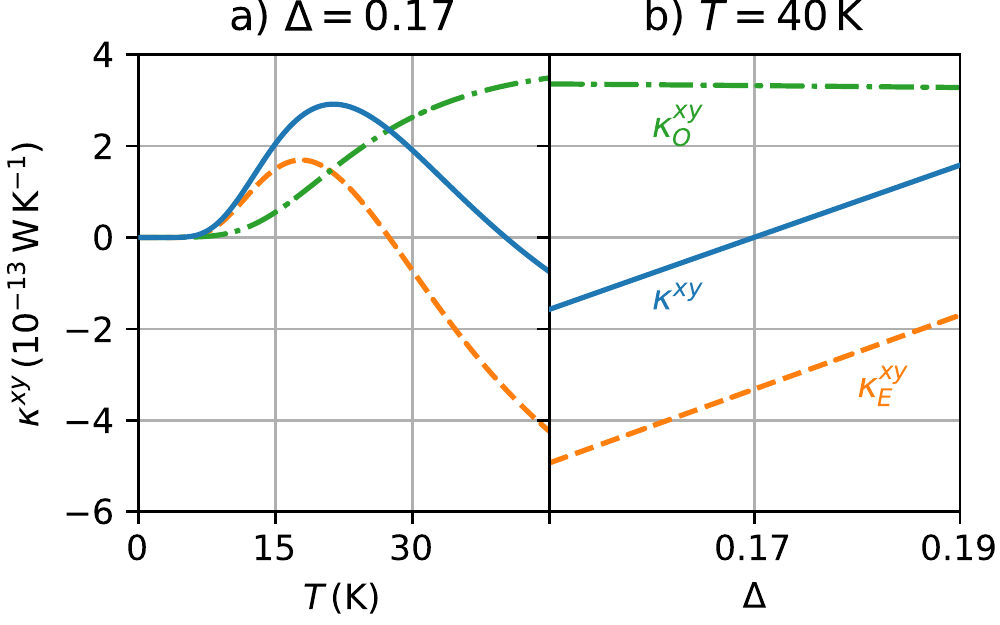}
    \caption{\small{} (Colour online) The total thermal conductivity ${\kappa^{xy}}$ (blue, solid) and its current ${\kappa^{xy}_E}$ (yellow, dashed) and orbital ${\kappa^{xy}_O}$ (green, dot-dashed) contributions against a) the temperature $T$ at ${\Delta=0.17}$ and b) the KOC parameter $\Delta$ at ${T = \SI{40}{K}}$ at the point in phase~III from Fig.~\ref{fig:kTphase}.}
    \label{fig:cntrb}
\end{figure}

\subsection{Phase II}

In this phase, only bands 3 and 4 have non-vanishing Chern numbers and provide significant contributions to ${\kappa^{xy}}$. Band 3 has ${C_3=1}$ and the higher energy band 4 with ${C_4=-1}$ can never fully counter it's negative contribution so that this is the only phase where the thermal Hall conductivity has a fixed sign with ${\kappa^{xy}<0}$. This is similar to the isotropic lattice, where the system is reduced to a two-band model with the upper (lower) band having a Chern number of ${-1(1)}$~\cite{Owerre2016c}. At the boundary with phase~III where the magnitude of $\kappa^{xy}$ is greatest, we have ${\Delta=0}$ so that the KOC modulation is absent and the system reduces to this two-band model exactly.

\subsection{Phase III}

The nonzero winding numbers in this phase are ${\nu_1=\nu_2=-\nu_3=\nu_4=\nu_5=-1}$, so we find four edge modes propagating counterclockwise against the fifth, in band gap 3, propagating clockwise. As shown in Fig.~\ref{fig:kT}c), at low temperatures the lower bulk bands dominates with ${\nu_1=\nu_2=-1}$ so that ${\kappa^{xy}>0}$. Upon increasing the temperature, the edge mode in band gap 3 becomes populated, with the sign change in ${\kappa^{xy}}$ occurring due to the transition to its dominance over the lower edge modes.

\begin{figure}[t]
    \centering
    \includegraphics[width=\linewidth]{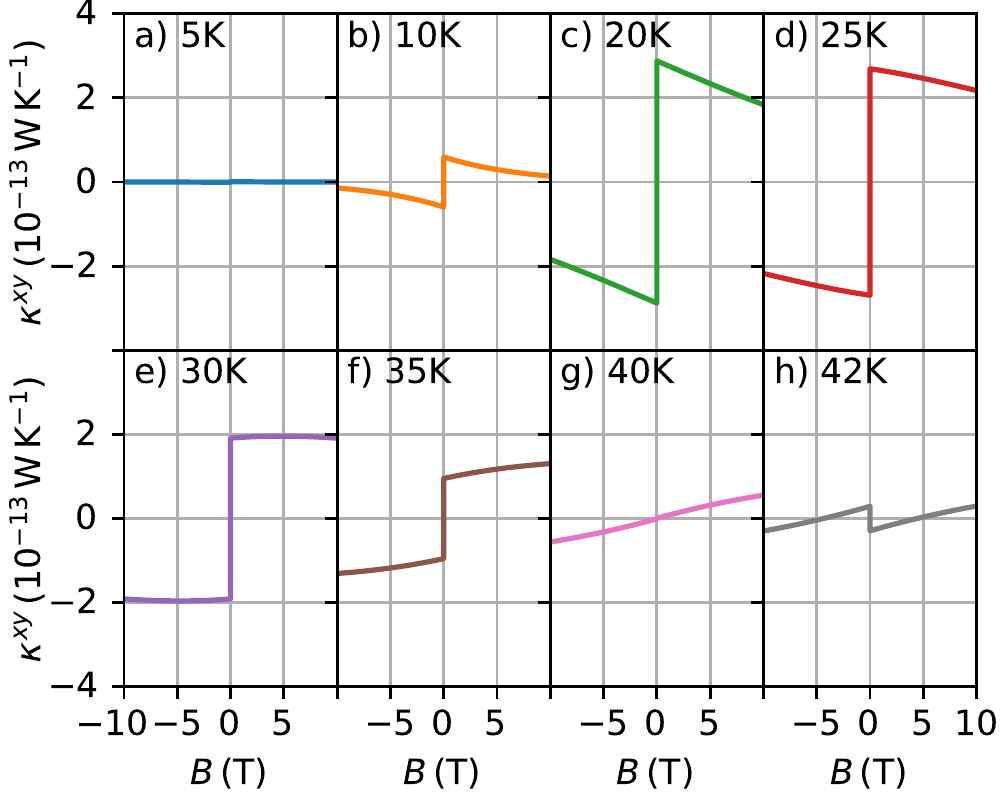}
    \caption{\small{} (Colour online) The thermal Hall conductivity ${\kappa^{xy}}$ against magnetic field $B$ at various temperatures $T$ at the point in phase~III in Fig. \ref{fig:kTphase}.}
    \label{fig:kB}
\end{figure}

On the other hand, the total thermal Hall conductivity ${\kappa^{xy}}$ in Eq.~(\ref{eq:kxy}) and its contributions due to the current ${\kappa^{xy}_E}$ in Eq.~(\ref{eq:kxyE}) and orbital motion of magnons ${\kappa^{xy}_O}$ at the point in phase~III with ${\Delta=0.17}$ and the other parameters from Fig.~\ref{fig:kTphase} are plotted against temperature in Fig.~\ref{fig:cntrb}a). This lies on the line of points throughout the parameter space where ${\kappa^{xy}}$ is vanishing at ${T = \SI{40}{K}}$. At low temperatures, ${\kappa^{xy}_E>0}$ dominates and ${\kappa^{xy}}$ rapidly increases to its peak value of ${\SI{2.9e-13}{\watt\per\kelvin}}$ at ${T=\SI{21}{K}}$ in Fig.~\ref{fig:cntrb}a). Beyond this, ${\kappa^{xy}_E}$ rapidly decreases and, despite the large ${\kappa^{xy}_O>0}$, ${\kappa^{xy}}$ transitions to negative values at ${T=\SI{40}{K}}$. Thus, externally varying the temperature in the region around ${\SI{40}{K}}$ would allow direct control over the sign of ${\kappa^{xy}}$. The variation of these contributions to ${\kappa^{xy}}$ with respect to $\Delta$ around its critical value of 0.17 at ${\SI{40}{K}}$ is depicted in Fig.~ \ref{fig:cntrb}b). Since the orbital contribution ${\kappa^{xy}_O}$ is nearly constant around criticality in both cases, this demonstrates that the sign change is driven by the current contribution ${\kappa^{xy}_E}$ at this point. Alternatively, as shown in Fig. \ref{fig:kT}c), the sign change of ${\kappa^{xy}}$ with temperature is a consequence of a competition between bands of different Chern numbers. For a given temperature, as shown in Fig. \ref{fig:kTphase}, the sign change in ${\kappa^{xy}}$ is due to a competition between the contributions ${\mathcal{D} \kappa^{xy}_\lambda}$ of isoenergy surfaces near the $\Gamma$ and $K$ points, as in Sec.~\ref{ssec:I}.

\subsection{Phase IV}

Similarly to phases I and III, in this phase ${\kappa^{xy}}$ changes its sign in the parameter space at fixed temperature. However, as shown in Fig. \ref{fig:kTphase}, this sign change occurs for small values of ${\Delta}$ and ${D}$. For larger values, ${\kappa^{xy}}$ is positive. The set of Chern numbers for this phase is ${(-1,0,1,-1,0,1)}$ with winding numbers ${\nu_1=\nu_2=\nu_4=\nu_5=-1}$ and $\nu_3=0$, so we find four edge modes propagating counterclockwise. Thus, we generally have ${\kappa^{xy}>0}$ in this phase. Alternatively, for the point given in Fig.~\ref{fig:kTphase} at low temperatures, band 1 with ${C_1=-1}$ dominates, resulting in ${\kappa^{xy}>0}$. As the temperature increases, there is a competition of higher bulk bands, however their contribution to the Hall current is strongly suppressed by the $c_2$ function so that ${\kappa^{xy}}$ remains positive.

\subsection{\label{ssec:B}Magnetic Field}

In the Hall effect of electrons, time reversal symmetry is broken as a result of an external magnetic field $B$ deflecting the electrons carrying the heat currents via the Lorentz force. In the thermal Hall effect of magnons, which are electrically neutral, a similar effect is a consequence of the DMI and does not require an external magnetic field. Thus, in a real ferromagnetic material, the observation of the Hall effect at weak $B$ would directly demonstrate that the particles carrying the heat currents are electrically neutral, being either magnons or phonons~\cite{Onose2010}. As discussed in Sec.~\ref{sec:model}, $B$ affects the magnon bands through the Zeeman term in Eq.~(\ref{eq:H}). Flipping the sign of $B$ effectively flips the sign of the DMI strengths, so that ${\kappa^{xy}(-B)=-\kappa^{xy}(B)}$. Increasing $|B|$ or the easy-axis anisotropy $A$ increases the on-site potential $\epsilon_0$ and the energy gap $h$ between the ground state and band 1 at $\Gamma$, shown in Fig.~\ref{fig:iso}. This suppresses the contributions to the thermal Hall conductivity ${\kappa^{xy}_\lambda}$ of all bands as a consequence of the $c_2$ function in Eq.~(\ref{eq:kxy}).

Plots of the thermal Hall conductivity ${\kappa^{xy}}$ against the magnetic field $B$ for various temperatures $T$ at the point in phase~III from Fig.~\ref{fig:kTphase} are depicted in Fig.~\ref{fig:kB}. We now consider ${B>0}$ for simplicity. For low temperatures, the occupation of all bands are so strongly suppressed that ${\kappa^{xy}}$ is negligible, as shown for ${T = \SI{5}{K}}$ in Fig.~\ref{fig:kB}a). Increasing the temperature with ${T < \SI{20}{K}}$, only band 1 has a significant contribution ${\kappa^{xy}_1 > 0}$, so increasing $B$ decreases ${\kappa^{xy}}$. This continues as band 3 with ${\kappa^{xy}_3 < 0}$ becomes successively occupied up to ${\SI{30}{K}}$, at which point both bands are suppressed equally and ${\kappa^{xy}}$ is nearly invariant with respect to increases in $B$. Above this temperature, as band 3 dominates and decreases ${\kappa^{xy}}$, increasing $B$ increases ${\kappa^{xy}}$. At ${T = \SI{40}{K}}$, when ${\kappa^{xy}=0}$ at ${B=0}$, we find no discontinuity in the plot as shown in Fig.~\ref{fig:kB}f). Above this, we have ${\kappa^{xy}<0}$ so that increasing $B$ increases ${\kappa^{xy}}$. This results in a transition to positive values so that it is possible to control the sign of ${\kappa^{xy}}$ beyond the simple transition at ${B=0}$ as shown for ${T = \SI{42}{K}}$ in Fig.~\ref{fig:kB}g). Observation of this suppression of ${\kappa^{xy}}$ by increase $|B|$ would preclude the heat currents being carried by phonons, as their mean free path and hence ${\kappa^{xy}}$ is expected to be enhanced due to reduced magnon-phonon scattering~\cite{Onose2010}.

\section{\label{sec:conc}Conclusion}

We have investigated the thermal Hall effect of magnons in a KOC modulated honeycomb ferromagnetic lattice with DMI. As an extension of our previous model \cite{Pantaleon2018b}, we have considered a strain based mechanism for introducing this modulation. We have calculated the thermal Hall conductivity ${\kappa^{xy}}$ at low temperature in the four different topological phases as a function of the KOC parameter, DMI strength, temperature and external magnetic field. We found that in the topologically trivial phase, where all the Chern numbers are vanishing, the thermal Hall conductivity is nonzero due to a non-vanishing local Berry curvature emerging from band proximity effects. We also found that ${\kappa^{xy}}$ can easily be controlled using the external temperature and magnetic field as well as the internal KOC and DMI parameters. Further, we found that the sign of ${\kappa^{xy}}$ is not fixed, exhibiting sign changes with respect to these parameters. Thus, such a material with both KOC modulation and DMI would be highly appropriate for thermal components.

Finally, near the Curie temperature, magnon-magnon interactions beyond the scope of the linear spin-wave theory considered become important~\cite{Poling1982}. Furthermore, to derive the Grunessein-like parameter of ${\mathrm{CrI}_3}$ in a similar manner to graphene~\cite{Cheng2011}, it would be necessary to consider other interactions such as magnon-phonon. We will report on the results of such investigations in the future.

\section*{Acknowledgments}

We acknowledge useful discussions with Elias Andrade and Rory Brown. Pierre A. Pantale{\'o}n is sponsored by Mexico's National Council of Science and Technology (CONACYT) under scholarship No. 381939. P.A.P and R.C-B acknowledge Programa de Movilidad Acad{\'e}mica 2018 UABC. 

\bibliographystyle{apsrev4-1}

\bibliography{report,soc}
\end{document}